\documentclass[12pt]{article}
\usepackage{latexsym,amsfonts,amsmath,amsthm,amssymb}

\title{Bell's inequality for conditional probabilities and
nonexistence of a realistic model for the two dimensional Hilbert space}

\author{Andrei Khrennikov\footnote{Supported in part by the EU Human
Potential Programme, contact HPRN--CT--2002--00279 (Network on
Quantum Probability and Applications) and Profile Math. Modelling
in Physics and Cogn. Sc. of V\"axj\"o University.} \\International Center for Mathematical
Modeling \\ in Physics and Cognitive Sciences,\\
University of V\"axj\"o, S-35195, Sweden\\
Email:Andrei.Khrennikov@msi.vxu.se}

\date{}
\begin{document}
\maketitle

\begin{abstract} We perform the analysis of probabilistic assumptions of
Bell's approach. We emphasize that J. Bell wrote about probability without to specify the concrete
axiomatics of probability theory. The careful analysis demonstrated that (surprisingly) J. Bell did not apply the classical probability model (Kolmogorov) to describe ``classical physical framework.'' In fact, he created his own probabilistic model and compared it with the quantum one.
The crucial point is that J. Bell did not pay attention to {\it conditional probabilities.} We show   that conditional probability in his model cannot be defined by classical Bayes' formula.  We also use the approach based on Bell-type inequalities in the conventional probabilistic approach, Kolmogorov model. We prove an analog of Wigner's
inequality  for conditional probabilities and by using this inequality show that
predictions of the conventional and quantum probability models disagree already in the case of noncomposite systems
(even in the two dimensional case!).
\end{abstract}

Bell's inequality has  definitely beaten all
records in the history of physics on the number of possible
interpretations, misunderstandings, mysteries and revolutionary
consequences. As the organizer of a series of international
conferences on foundations of quantum theory, see e.g. [1]--[5] , I was
really disappointed by stormy debates on Bell's inequality. There is
no any problem in quantum foundations which can be compared with
Bell's inequality by intensiveness  of discussions and reactions to
opponent's views.

I asked myself many times: ``What is so special in
Bell's inequality?'' I concluded that the main problem was that from
the very beginning J. Bell formulated precisely the aim of his
investigation, namely, to prove nonlocality of QM.\footnote{People
who were very close to J. Bell explained me that he was ``nonlocal
realist'' and the main aim his investigations was to find arguments
supporting the Bohmian model of QM.}. But he did not determine
precisely  a mathematical model which he used to formalize the
problem. J. Bell did not define precisely the probabilistic rules of the game
in the form of the axiomatic probability model which he planed to apply to
investigation of the EPR-Bohm experiment. Therefore different people presented
their own views on mathematical rules of the Bell-game and,
consequently, they come to totally different conclusions.

J. Bell wrote about probability [6] without to specify the concrete
axiomatics of probability theory. Unfortunately, after numerous
conversations with outstanding physicists I have the impression that
there is a rather common opinion that there is no need in using of
an axiomatic model for probability. Typically such a viewpoint is
motivated by by considering probability as ``physically well defined
quantity.'' Therefore one need not take care of probabilistic formalism.
I totally disagree with such a viewpoint. In the same way one could
say that physicist need not take care of a mathematical model of
space. Fortunately, nobody would say this now days. Everybody
understood that it is one thing to work in the continuous real space
(as people do in classical physics) and totally different thing to
work in discrete or $p$-adic space . Nobody would say that physicist
need not take care of geometry. Suppose that people would try to
work in physics without to determine mathematical model of space and
geometry on it. It is clear that such an activity would induce
debates and misunderstandings and variety of paradoxes. I think that
the similar thing happened with Bell's inequality (but with
probability, instead of space).

Nevertheless,  during last 40 years people have been trying to
proceed with Bell's inequalities without to describe precisely the probabilistic model.
Moreover, there is a rather common opinion that  it is possible
to work only with  frequencies and hence there is no need in the
mathematical description of a corresponding probability model.

In this paper we describe precisely the Bellian probabilistic model which was really used in the EPR-Bell
framework. It may be surprising that in fact J. Bell did not apply the classical probabilistic
model, the Kolmogorovian model  [7] (which was used in the classical statistical mechanics).
The Bellian classical probabilistic model, see section 1.1, differs crucially from the conventional Kolmogorovian model of probability. So J. Bell did not compare quantum model (based on the Hilbert space calculus of probabilities)
with the conventional Kolmogorovian model, but he created (without to pay attention to this) his own
``classical probability model´´ and compared the latter with quantum mechanics. But the the classical statistical mechanics is described by the Kolmogorov model, and not by Bellian model. Therefore already on the `classical level''
J. Bell lost direct connection with classical physics. He used a rather unusual ``classical probability model´´
and it is not so strange that he came to rather unusual explanation of difference between it and quantum mechanics.

The crucial point is that J. Bell did not pay attention to {\it conditional probabilities.} But in any well established
probability model conditional probabilities should be defined, cf. [8], [9]. In the Kolmogorov model they are defined by the Bayes'
formula, in the quantum model by von Neumann projection postulate. But in the Bellian ``classical probability model''
there is no conditional probabilities! The only thing that evident from Bell's works is that conditional probability in his model cannot be defined by Bayes' formula. We show this in the present note. We also use the approach based on Bell-type inequalities in the conventional probabilistic approach, Kolmogorov model. We prove an analog of Wigner's
inequality [10] (a Bell-type inequality) for conditional probabilities and by using this inequality show that
predictions of the conventional and quantum probability models disagree already in the case of noncomposite systems(even in the two dimensional case!).

\section{Measure-theoretical derivation of Bell's inequality}
\subsection{Bellian probability model}
Let us consider a probabilistic model which based on the measure-theoretic approach (as the original Kolmogorov probability model), but there is no such a notion as conditional probability. In particular, there is no Bayes' formula which defines conditional probability in the Kolmogorov model. Such a probability model we call {\it Bellian model,}
since J. Bell was the first who started to work in such a framework: measure-theoretical approach without Bayes formula and, in fact, without any definition of conditional probability.

\subsection{Bell's inequality}

Let  ${\cal P}=(\Omega, {\cal F}, {\bf P})$ be a Kolmogorov probability space.
The Bell's inequality [6] can be mathematically formulated in the form of the following theorem.

{\bf Theorem 1.1.} (Bell inequality for covariations) {\it Let
$a,b,c= \pm 1$ be random variables on ${\cal P}.$  Then Bell's
inequality
\begin{equation}
\label{BBB} \vert <a,b > - < c,b >\vert \leq 1 - <a,c>
\end{equation}
holds true.}

\section{Bell-Wigner inequality for conditional probabilities and
measurements of spin projections for a single electron}
\subsection{Statistical model of classical physics}

{\bf Definition 2.1.} {\it Let ${\cal O}= \{a, b, \ldots c\}$ be a
system of physical observables. This system permits the
classical statistical description if there
exists a Kolmogorov probability space ${\cal P}=(\Omega, {\cal F},
{\bf P})$ such that all observables belonging to ${\cal O}$ can be
represented by random variables on ${\cal P}$ and conditional
probabilities  are defined by the Bayes' formula:
${\bf P}(a=\alpha/b=\beta)=\frac{{\bf P}(a=\alpha, b=\beta)}{ {\bf P}(b= \beta)}.$}

Thus the possibility of the classical statistical description of
some system of physical observables is equivalent to the possibility
of using the Kolmogorov probability model. We remark that classical statistical mechanics permits the classical
statistical description. We would like to analyze the possibility to
apply the  classical statistical description to quantum mechanics.
The Kolmogorov space ${\cal P}$ can be considered as a space of
hidden variables. Traditionally there is used the symbol $\Lambda,$
instead of $\Omega.$

\subsection{Wigner inequality}

We recall the following simple mathematical result, see Wigner [10]:

{\bf Theorem 2.1.} (Wigner inequality) {\it{Let $a, b, c=\pm 1$ be
arbitrary random variables on a Kolmogorov space ${\cal P}.$ Then
the following inequality holds true.}}

\begin{equation}
\label{BB} {\bf P} (a=+1, b=+1) + {\bf P}(b=-1, c=+1) \geq {\bf
P}(a=+1, c=+1)
\end{equation}

The proof of this theorem in the purely mathematical framework can
be found e.g., in my book [11], p. 89-90.\footnote{The inequality (\ref{BB})
is, in fact, the well known version of Bell's inequality obtained by
Wigner in 1970, see [10]. Moreover, Wigner proved this inequality in
the same general probabilistic framework as was used in Theorem 2.1
(so by using the Bellian probability model). Then following Bell's
strategy  he applied (\ref{BB}) to the EPR-Bohm experiment for
correlated particles. It is easy to see that (\ref{BB}) is violated
for an appropriative choice of pairs of spin projectors.}

\subsection{Wigner-type inequality for conditional probabilities}
 As a simple consequence of Theorem 2.1, we obtain:

{\bf Theorem 2.2.} (Wigner inequality for conditional
probabilities). {\it{Let $a, b, c=\pm 1$ be symmetrically
distributed observables which permit the classical statistical
description. Then the following inequality holds true:}}
\begin{equation}
\label{BB1} {\bf P}(a=+1/b=+1) + {\bf P}(c=+1/b=-1) \geq {\bf
P}(a=+1/c=+1)
\end{equation}

{\bf Proof.} We have ${\bf P}(b=+1)={\bf P}(b=-1) = {\bf
P}(a=+1)={\bf P}(a=-1)={\bf P}(c=+1)={\bf P}(c=-1)=1/2.$ Thus
$$
{\bf P}(a=+1/b=+1) + {\bf P}(c=+1/b=-1)
$$
$$
=2{\bf P}(a=+1, b=+1) + 2{\bf
P}(c=+1, b=-1)
$$
and $ {\bf P}(a=+1/c=+1)=2{\bf P}(a=+1, c=+1).$ Hence by (\ref{BB})
we get (\ref{BB1}).

We underline again that the main distinguishing feature of
(\ref{BB1}) is the presence of {\it only conditional probabilities.}
Conditional probabilities can always be calculated by using quantum
formalism. In fact, we need not consider pairs of particles, since
conditional probabilities are well defined even for noncomposite quantum
systems.

\subsection{The impossibility of classical statistical description of
spin projections of a single electron}

Suppose  that the classical
statistical description can be used for spin $\frac{1}{2}$ system.
Thus all spin projections can be represented by random variables on
a single Kolmogorov space $\cal P$ and (in the opposite to the
Bellian probability model) conditional probabilities are given by
the Bayes' formula.

We consider a family of spin projections: $\sigma(\theta)=\cos
\theta \sigma_z + \sin\theta \sigma_x,$ where $\sigma_x, \sigma_z$
are Pauli matrices, $\theta \in [0, 2\pi).$

The classical (Kolmogorovian) statistical realism implies that we
can apply the inequality (\ref{BB1}) to any three spin projectors
$\sigma(\theta_1), \sigma(\theta_2), \sigma(\theta_3)$:
\begin{equation}
\label{BB2} {\bf P}(\sigma(\theta_1)= +1/\sigma(\theta_2)= +1) +
{\bf P}(\sigma(\theta_3)= +1/\sigma(\theta_2)= -1)
\end{equation}
$$
\geq {\bf P}(\sigma(\theta_1)= +1/\sigma(\theta_3)= +1) .
$$

As always we can compute conditional probabilities by using quantum
formalism [12]. If we have two arbitrary dichotomous
observables $a$ and $b$ which can be described by QM then (see chapter 1, section 6):
$$
{\bf P}(a=\alpha_i/b=\beta_j)=|<e_i^a, e_j^b>|^2,
$$
where $\{e_i^a\}$ and $\{e_i^b\}$ are systems of normalized
eigenvectors of operators $\hat{a}$ and $\hat{b},$ respectively: $
\hat{a} e_i^a=\alpha_i e_i^a, \hat{b} e_j^b=\beta_j e_j^b.$ For spin
projectors we have
$$
\sigma(\theta) \varphi_+(\theta)=\varphi_+(\theta), \;
\mbox{where}\; \varphi_+(\theta)=(\cos \frac{\theta}{2}, \sin
\frac{\theta}{2})
$$
$$
\sigma(\theta) \varphi_-(\theta)=-\varphi_-(\theta), \;
\mbox{where}\; \varphi_-(\theta)=(-\sin \frac{\theta}{2}, \cos
\frac{\theta}{2})
$$

Thus
$$
{\bf P}(\sigma(\theta_1)= +1 /\sigma(\theta_2)=+1)=\cos^2
\frac{\theta_1 - \theta_2}{2},
$$
$$
{\bf P}(\sigma(\theta_3)= +1 /\sigma(\theta_2)=-1)=\sin^2
\frac{\theta_3 - \theta_2}{2},
$$
$$
{\bf P}(\sigma(\theta_1)= +1 /\sigma(\theta_3)=+1)=\cos^2
\frac{\theta_1 - \theta_3}{2}.
$$

By (\ref{BB2}) we have
$$
\cos^2 \frac{\theta_1-\theta_2}{2} + \sin^2
\frac{\theta_3-\theta_2}{2} \geq \cos^2 \frac{\theta_1-\theta_3}{2}
$$
We take $\theta_1=0, \theta_2=6\theta, \theta_3=2\theta$ and we get
the following trigonometric inequality:

$$\cos^2 3\theta + \sin^2 2\theta \geq \cos^2 \theta.$$

The latter inequality  was applied by Wigner at the last step of his
analysis of Bell's arguments for the EPR-Bohm experiment with
correlated particles. It is well known [] that this trigonometric
inequality is violated for sufficiently large $\theta.$

\medskip

{\bf Conclusion.} {\it  A prequantum classical statistical model
(Kolomogorovian model) does not exist even for the Hilbert space of
the dimension ${\bf d=2}.$}\footnote{This result does not contradict to Bell's
example [6], since the conventional definition of the classical probability model
differs from that J. Bell had in mind (unfortunately, he did not present a precise mathematical
description of his classical probabilistic model). Our definition  coincides with that
one which is used in classical statistical mechanics. Thus from the
probabilistic viewpoint classical statistical mechanics differs from
quantum mechanics even for ensembles of noncomposite quantum systems.}

\section{Rules of correspondence between classical and quantum statistical models}

In the previous section we demonstrated that the classical probabilistic formalism (Kolmogorovian measure-theoretic model) is incompatible with the quantum probabilistic formalism (Dirac-von Neumann complex Hilbert space model) already in the case of noncomposite quantum systems. By using the generalization of Wigner inequality to conditional probabilities, Theorem 2.2, we have shown that classical conditional probabilities (defined via Bayes' formula, see
section 1 of chapter 2) are incompatible with quantum conditional probabilities (defined via Hilbert space projections of states. Personally I thing that this would be the correct application of Bell-type inequalities in quantum physics. The main mystery of quantum physics is not nonlocality  or ``death of reality'',
but the violation of Bayes' rule for quantum conditional probabilities.

In fact, the definition of quantum conditional probabilities is based  on the von Neumann projection postulate. Thus the main mystery of quantum theory is encoded in the von Neumann projection postulate.

Unfortunately, J. Bell did not choose such a strategy from the very beginning. Moreover, he did not pay attention to the  problem of correspondence between classical and quantum conditional probabilities.   J. Bell chosen a totally different startegy. He introduced into consideration a new element namely nonlocality. However,
it is clear that by demonstrating the violation of Bell-type inequalities for composite systems one could not clarify the violation of the conditional probabilistic version of Wigner's inequality, (\ref{BB1}), already for noncomposite systems.

We say a few words about the mathematical definition of realism  in discussion on completeness of quantum mechanics. It is assumed that there exists a space of hidden variables  $\Omega$
 representing states of individual physical systems. On this space there are defined
objective physical variables, $\xi: \Omega \to {\bf R}.$ Denote the space of physical variables by $V(\Omega)$
(this is some space of real-valued functions on $\Omega).$\footnote{The choice of this functional space depends on a model.}
On the other hand, there is considered a space of physical observables $O$. In the quantum model they are represented by self-adjoint operators. The whole HV-story is about the correspondence between the space of physical variables
$V(\Omega)$ and the space of physical observables $O;$  about the possibility to construct a map
\begin{equation}
\label{MP}
T: V(\Omega)\to O .
\end{equation}
The main problem is that still nobody knows precisely which features such a map $T$ should have.
We recall the history of this problem. J. von Neumann was the first who presented a list of possible features of $T,$
see [12].
His main postulates on the classical $\to$ quantum correspondence were:

\medskip

VN1). $T$ is one-to-one map.\footnote{Different physical variables induce different physical observables and any physical observable correspond to some physical variable; so the map $T$ is surjective: $T(V(\Omega))= O.$ We pay attention that the latter feature of $T$ played the fiundamental role in von Neumann's considerations, [12].}

\medskip

VN2). For any Borel function $f: {\bf R} \to {\bf R},$ we have $T(f(\xi)) = f(T(\xi)).$

\medskip

VN3). $T(\xi_1+ \xi_2+ ...) = T(\xi_1) + T(\xi_2)+...$ for any any sequence $\xi_k \in V(\Omega).$\footnote{As J. von Neumann remarked: ``the simulateneous measurability of $T(\xi_1), T(\xi_2), ...$ is not assumed'', see [12], p. 314.}

\medskip

VN4). The range of values  of the observable $T(\xi)$ coincides with the range of values of the variable $\xi.$

\medskip
We remark that any statistical model should also contain a space of statistical states.
In a prequantum statistical model statistical states are represented by measures on $\Omega,$ denote this space by
$S_{\rm{class}}(\Omega);$ in the quantum model -- by von Neumann density operators; denote this space $S_{\rm{quant}}.$
It was assumed that

\medskip

VN5). The correspondence
$$
t: S_{\rm{class}}(\Omega) \to S_{\rm{quant}}
$$
is one-to-one.

\medskip

VN6). Assume that for $\xi_1,..., \xi_n\in V(\Omega)$ the corresponding observables
$T(\xi_1),..., T(\xi_n)$ can be measured simultaneously. Then
the simultaneous probability distributions of the variables $\xi_1,..., \xi_n$ (with respect to a statistical state $\rho \in S_{\rm{class}}(\Omega))$ and the observables $T(\xi_1),..., T(\xi_n)$ (with respect to a statistical state $D=t(\rho) \in S_{\rm{quant}})$ coincide.

\medskip

J. von Neumann did not write  anything about conditional probability.  But it should be natural to complete his system of postulates by

\medskip

CP). For any pair of physical variables $\xi$ and $\eta$ and a statistical state $\rho\in S_{\rm{class}}(\Omega),$   the  probability (given by the Bayes' formula) that
$\xi\in B $ under the condition $\eta\in A$  coincides with
the probability (given by the quantum formula for conditional probability) to observe that $T(\xi)\in B$ after a measurement of $T(\eta)$ on the state $D=t(\rho)\in S_{\rm{quant}}$ such that $T(\eta)\in A.$

Here $A$ and  $B$ are Borel subsets of the real line.

\medskip

J. von Neumann proved that a correspondence $T$ between a classical statistical model $M=(S_{\rm{class}}(\Omega), V(\Omega))$ and the quantum statistical model $N=(S_{\rm{quant}}, O)$ satisfying the postulates   VN1--VN6 does not exist.  J. Bell critisized the von Neumann postulates  VN1, VN3  as unphysical (we remark that Kochen and Specker also used the postulate VN1).  He deleted VN3 from the list of postulates on the correspondence between $M$ and $N$ and modified VN1 in the following way:

\medskip

VN1B). $T$ maps $V(\Omega)$ onto $O.$\footnote{Any quantum observable $a\in O$ has a preimage (may be nonunique) in
$V(\Omega).$}

\medskip

The crucial point was that neither J. Bell nor J. von Neumann paid attention that without
the postulate CP on the correspondence between classical and quantum condition probabilities
such a correspondence would look very unnatural. In the opposite to the postulates VN1, VN3, this postulate does not look so unphysical (especially in combination with other postulates).
One should explain why the Bayes' formula which worked well in classical statistical physics does not work in quantum physics. And consideration of correlated particles could not clarify anything.

\section{Connection with papers of Accardi, Aerts, Pitowsky, Czachor}

In his Email-comment to the second version of this preprint.
M. Czachor paid my attention to his paper [13] on
classical models of spin which contains an extended analysis of models of I. Pitowsky [14] and Aerts [15].
I totally agree with conclusions of M. Czachor that ``both these hidden variable models are based on an observation that a structure of conditional probabilities characteristic for systems with spin is not a Kolmogorovian one. The problem is rooted
in a non-Bayesian structure of such probabilities and is typically manifested by a violation of Bell's inequality.''

Especially interesting for us is M. Czachor's investigation to find out which elements of models are sufficient
for the non-Kolmogorovity of description and his conclusion that ``a conditioning by a change of state is,
actually, the required sufficient condition.''

I also think that both Pitowsky's and Aerts' models are not about simultaneous measurements as we have in the
EPR-Bohm framework, but about conditional measurements:

{\small We have some state $\phi;$ perform a measurement of an observable $a;$ the state $\phi$ is changed;
in a new post $a$-measurement state we perform a measurement of another observable $b$ which is incompatible
with (or better to say complementary to)  $a.$}

Of course, both Pitowsky's and Aerts' models were created in relation to Bell's inequality. And M. Czachor
pointed out that ``There is no contradiction with the Bell Theorem, because it is impossible to derive the Bell
inequality for this model.''

I agree that it is impossible to get the ordinary Bell's inequality, because we cannot perform a simultaneous
measurement of $a$ and $b$. This observation was the starting point for the investigation presented in my paper.
Instead of Bell's inequality for the simultaneous probability distributions, I derived Bell's inequality
for {\it conditional probabilities.} This inequality can be applied to conditional measurements. I demonstrated
that it is violated by quantum model (as everybody could expect!). We remark that this conditional probability
inequality is based only on the assumption that we can use Bayes' formula for conditional probabilities, cf.
L. Accardi [16], [17]. Since both Pitowsky's and Aerts' models reproduce quantum probabilities, Bell's inequality for
conditional probabilities is automatically violated for these models.

I would like to thank L. Ballentine, P. Busch,  S. Gudder, B. Coecke,
W. De Baere, W. De Muynck for discussions on the statistical structure of quantum theory.

{\bf References}

1. Proceedings of Conference
{\it Foundations of Probability and Physics,} Ser.  Quantum Probability and White Noise Analysis,
{\bf 13}, 201- 218, WSP, Singapore, 2001.

2. Proceedings of Conference {\it Quantum Theory: Reconsideration
of Foundations.} Ser. Math. Modelling in Phys., Engin., and Cogn. Sc., v.2,
V\"axj\"o Univ. Press, 2002.

3. Proceedings of Conference
{\it Foundations of Probability and Physics-2,}
Ser. Math. Modelling in Phys., Engin., and Cogn. Sc., v.5,
V\"axj\"o Univ. Press, 2003.

4. Proceedings of Conference {\it Quantum Theory: Reconsideration
of Foundations-2,} Ser. Math. Modelling in Phys., Engin., and Cogn. Sc., v. 10,
V\"axj\"o Univ. Press, 2004.

5. Proceedings of Conference
{\it Foundations of Probability and Physics-2,}
American Institute of Physics, 2005.

6. J. S. Bell, {\it Speakable and unspeakable in quantum mechanics.}
Cambridge Univ. Press (1987).

7. A. N. Kolmogoroff, {\it Grundbegriffe der Wahrscheinlichkeitsrechnung.}
Springer Verlag, Berlin (1933); reprinted:
{\it Foundations of the Probability Theory}.
Chelsea Publ. Comp., New York (1956);

8. S. P. Gudder, Trans. AMS 119 (1965) 428.

S. P. Gudder, Axiomatic quantum mechanics and generalized probability theory,
Academic Press, New York, 1970.

S. P. Gudder, An approach to quantum probability,
in:  A. Yu. Khrennikov (Ed.),
Foundations of Probability and Physics, Q. Prob. White Noise Anal. 13,  WSP, Singapore, 2001, p. 147.

9.  L. E. Ballentine,  Rev. Mod. Phys. 42 (1970) 358.

L. E. Ballentine, Quantum mechanics, Englewood Cliffs,
New Jersey, 1989.

L. E. Ballentine,  Interpretations of probability and quantum theory,
in:  A. Yu. Khrennikov (Ed.),
Foundations of Probability and Physics, Q. Prob. White Noise Anal. 13,  WSP, Singapore, 2001, p. 71.

10. E. P. Wigner, On hidden variables and quantum mechanical probabilities.
{\it Am J. Phys.}, {\bf 38}, 1005 (1970).

11.  A. Yu. Khrennikov, Interpretations of Probability. VSP Int. Sc. Publishers, Utrecht/Tokyo, 1999.

12. J. von Neumann, Mathematical foundations
of quantum mechanics, Princeton Univ. Press, Princeton, N.J., 1955.

13. M. Czachor, On classical models of spin, quant-ph/0205010.

14. I. Pitowsky, {\it Phys. Rev.} D, {\bf 27}, 2316 (1983).

15. D. Aerts, {\it J. Math. Phys.}, {\bf 27}, 202 (1986).

16. L. Accardi, in {\em The wave--particle dualism.  A tribute to Louis de Broglie on his 90th
Birthday,} ed. S. Diner, D. Fargue, G. Lochak and F. Selleri
(Reidel, Dordrecht, 1984), pp. 297--330.

17. L. Accardi, Phys. Rep., {\bf 77}, 169(1981).

\end{document}